\documentclass[lettersize,journal]{IEEEtran}
\usepackage{amsmath,amsfonts}
\usepackage{algorithmic}
\usepackage{algorithm}
\usepackage{array}
\usepackage[caption=false,font=normalsize,labelfont=sf,textfont=sf]{subfig}
\usepackage{textcomp}
\usepackage{stfloats}
\usepackage{url}
\usepackage{verbatim}
\usepackage{graphicx}
\usepackage{cite}

\usepackage{hyperref}
\usepackage{xcolor}
\usepackage{adjustbox}
\usepackage{colortbl}
\usepackage{multicol}
\usepackage{multirow}
\usepackage{amssymb}
\usepackage{pifont}
\usepackage{caption}
\begin{document}

\title{LLM-Virus: Evolutionary Jailbreak Attack \\on Large Language Models}

\author{
    $\textbf{Miao Yu}^{1, *}$,
    $\textbf{Junfeng Fang}^{2, *}$,
    $\textbf{Yingjie Zhou}^{3}$,  
    $\textbf{Xing Fan}^{1, \dagger}$,
    $\textbf{Kun Wang}^{1, \dagger}$,
    $\textbf{Shirui Pan}^{4}$, 
    $\textbf{Qingsong Wen}^{1}$\\
    $^{1}$Squirrel Ai Learning\quad 
    $^{2}$National University of Singapore\\
    $^{3}$Sichuan University \quad
    $^{4}$Griffith University\\
    
    \thanks{Marker * denotes equal contributions, and $\dagger$ means that Xing Fan and Kun Wang are the corresponding authors.}
    \thanks{Correspondence to ymzgkxjsdx@mail.ustc.edu.cn}
}


\markboth{Journal of \LaTeX\ Class Files,~Vol.~14, No.~8, August~2021}%
{Shell \MakeLowercase{\textit{et al.}}: A Sample Article Using IEEEtran.cls for IEEE Journals}

\IEEEpubid{Perprint}

\maketitle

\begin{abstract}
While safety-aligned large language models (LLMs) are increasingly used as the cornerstone for powerful systems such as multi-agent frameworks to solve complex real-world problems, they still suffer from potential adversarial queries, such as jailbreak attacks, which attempt to induce harmful content. Researching attack methods allows us to better understand the limitations of LLM and make trade-offs between helpfulness and safety. However, existing jailbreak attacks are primarily based on opaque optimization techniques (e.g. token-level gradient descent) and heuristic search methods like LLM refinement, which fall short in terms of transparency, transferability, and computational cost. In light of these limitations, we draw inspiration from the evolution and infection processes of biological viruses and propose LLM-Virus, a jailbreak attack method based on evolutionary algorithm, termed evolutionary jailbreak. LLM-Virus treats jailbreak attacks as both an evolutionary and transfer learning problem, utilizing LLMs as heuristic evolutionary operators to ensure high attack efficiency, transferability, and low time cost. Our experimental results on multiple safety benchmarks show that LLM-Virus achieves competitive or even superior performance compared to existing attack methods. Our code is available at \href{https://github.com/Ymm-cll/LLM-Virus}{\textcolor{blue}{https://github.com/Ymm-cll/LLM-Virus}}.
\end{abstract}


\textcolor{red}{Warning: This paper contains potentially harmful text.}\\

\section{Introduction}
As LLMs emerge with exceptional and advanced capabilities such as knowledge \cite{sun2023head}, planning \cite{huang2024understanding} and reasoning \cite{yuan2024advancing}, they are exponentially being applied to systems (e.g. LLM-integrated applications \cite{greshake2023not} and LLM-based multi-agent systems \cite{yu2024netsafe}) across various domains and scenarios to solve certain complex problems. In this context, preventing the misuse of these powerful and influential LLM-based systems becomes increasingly critical \cite{liu2023trustworthy}. This research area, known as ``LLM Safety'', primarily focuses on preventing LLMs from being used for malicious behaviors, such as the spread of misinformation and bias, the generation of harmful content, and privacy breaches \cite{dong2024attacks, yao2024survey}. Directly issuing malicious queries is typically rejected, as most available LLMs (e.g. GPT and Claude) are safety-aligned via techniques like fine-tuning to ensure adherence of responses to secure human values \cite{ji2024beavertails, qi2023fine}. Unfortunately, a variety of jailbreak attack methods still exist that can bypass the built-in safety mechanisms \cite{mazeika2024harmbench}.

\begin{figure}[t]
    \centering
    \includegraphics[width=0.5\textwidth]{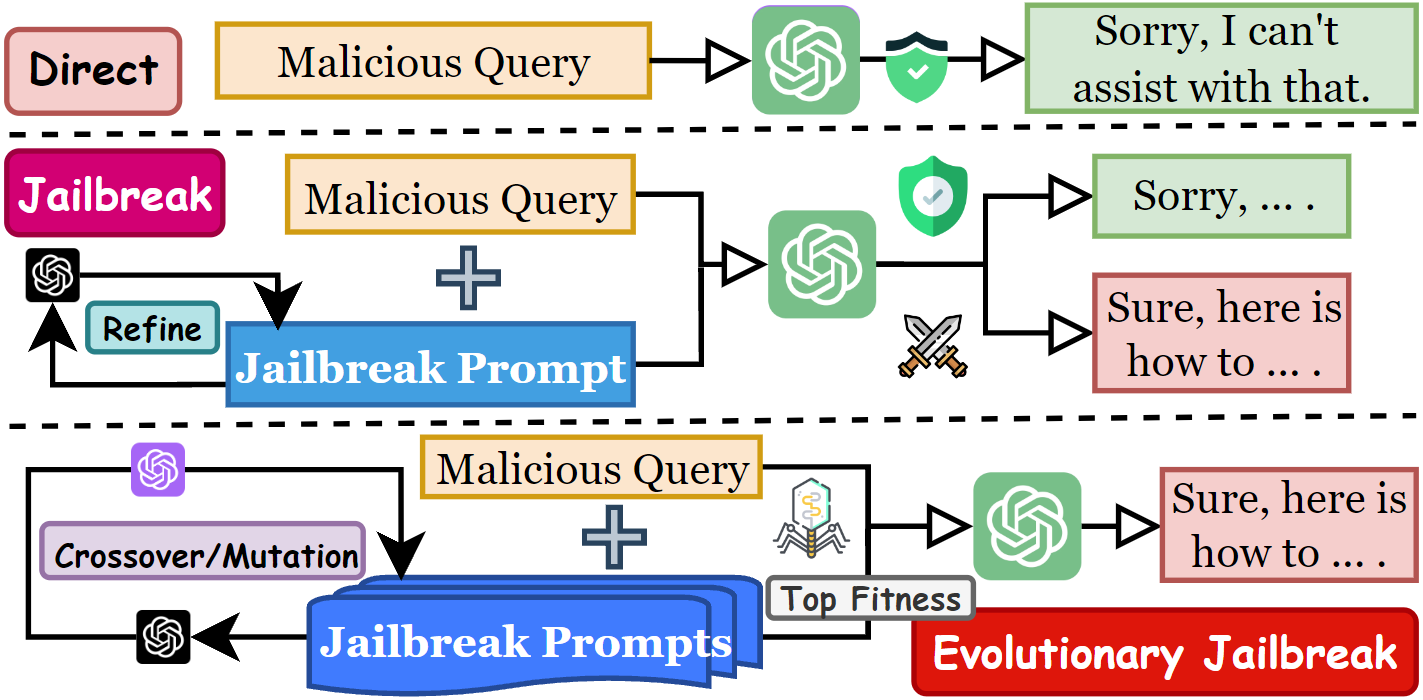}
    \vspace{-1.5em}
    \caption{Illustration of different ways for malicious querying (direct attack, normal jailbreak and evolutionary jailbreak).}
    \label{intro}
    \vspace{-1.5em}
\end{figure}

Attacks to LLM typically occur during training or inference \cite{dong2024attacks}, with the latter being more relevant to current usage scenarios, where users do not participate in the training process. Depending on whether the model is open-source, inference-time attacks can be categorized into two types \cite{yi2024jailbreak}: \textbf{White-box attacks} involve utilizing inherent information like gradient to optimize prefixes or suffixes that can elicit affirmative responses \cite{liu2023autodan, jones2023automatically}, which are then linked to malicious queries to induce desired responses \cite{zou2023universal}. \textbf{Black-box attacks}, in contrast, do not have access to the model's internal data and typically rely on manual techniques or LLM-generated methods to search for effective jailbreak prompts. For example, PAIR \cite{chao2023jailbreaking} leverages an attacker LLM to iteratively query a target LLM and refine the jailbreak prompt until successful.


\IEEEpubidadjcol
Additionally, a small number of studies have utilized evolutionary algorithms (EAs) to optimize attacks to models. (Figure \ref{intro}). For instance, AAA \cite{zeng2024ask} and \cite{gong2024cross} extends EAs to conduct evolutionary attacks in the image modality. As for those targeting at LLMs, AutoDAN \cite{liu2023autodan} optimizes existing jailbreak prompts by performing word-level mutations and paragraph-level crossover, with generation probability of affirmative prefix as fitness. BlackDAN \cite{wang2024blackdan} extends AutoDAN by applying the NSGA-II for multi-objective evolution. GPTFuzzer \cite{yu2023gptfuzzer} is inspired by EA-based fuzz testing to evolve jailbreak templates, while SMJ \cite{li2024semantic} focuses on improving the similarity between harmful actions and jailbreak prompts. Similar to existing works on LLM-enhanced EAs for combinatorial problems \cite{lange2024large, meyerson2023language, brownlee2023enhancing}, part of these methods use LLMs to aid mutation, crossover, or fitness evaluation to varying degrees.


However, there are several limitations in existing evolutionary jailbreak researches. \textbf{(I) Irrational Evolutionary Operator:} For instance, AutoDAN and BlackDAN only consider word-level mutation and random paragraph-level crossover. The former restricts the linguistic diversity, while the latter disrupts the contextual semantics of jailbreak prompt as a coherent piece of text. \textbf{(II) Limited Attack Scenario:} SMJ focuses solely on attacks against smaller, less secure models and only considers using LLM-based crossover by rephrasing. GPTFuzzer introduces additional LLM-based operators (e.g. expand and shorten), but it merely targets at toxicity. Besides, both of them conduct attacks to LLMs on a small malicious dataset (only 100 samples). \textbf{(III) High Time Cost \& Incomprehensive Baseline:} None of them address the increasing time cost associated with multiple iterations of evolution and enormous population size, nor do they provide a comprehensive performance comparison of evolutionary jailbreak with traditional (non-evolutionary) LLM attack methods.


To address the issues mentioned above, among others, we draw inspiration from infection and evolution of biological viruses and propose LLM-Virus, a black-box and efficient jailbreak attack based on LLM-enhanced EAs. Analogously, we treat the attack as viral infection. Jailbreak templates represent the mutating genetic material, while specific malicious queries as the functional proteins executing the attack. The LLM itself serves as the targeted host. Through selection driven by the host’s safety mechanisms, our goal is to evolve a population of virus strains (initialized by human-written jailbreak templates), with the help of LLMs as evolutionary operators.

Specifically, we first propose LLM-based crossover and mutation operations to explore wider solution space of jailbreak templates, encouraging specific textual properties such as diversity and conciseness. We treat jailbreak template as an individual to allow for the embedding of other queries within the template (thus more transferable), rather than a specific jailbreak prompt for one given malicious query. Then, similar to how viruses transfer and infect different hosts, we view the problem as a transfer learning task and introduce Local Evolution and Generalized Infection techniques to reduce computational and time costs while enhancing transferability and generalization. Furthermore, we employ more stringent attack classifiers to evaluate fitness (attack success rate) and perform experiments on HarmBench \cite{mazeika2024harmbench} and AdvBench \cite{zou2023universal} datasets. Our experiments comprehensively compare performance with traditional LLM attack methods, achieving state-of-the-art success attack rates and lower time costs. Moreover, we analyze the evolution dynamic of the average population fitness and perform ablation studies to demonstrate the effectiveness and validity of certain methods in LLM-Virus.


To conclude, our contributions are summarized as follows:
\begin{itemize}
    \item[\ding{182}] \textbf{\textit{Evolutionary Jailbreak:}} We propose LLM-Virus, an LLM attack via evolutionary algorithm, achieving extraordinary performance on multiple safety benchmarks.
    \item[\ding{183}] \textbf{\textit{New Insights:}} We treat the jailbreak evolution process as a transfer learning problem to optimize time consumption and transferability, with LLMs as evolutionary operators.
    \item[\ding{184}] \textbf{\textit{Holistic Experiments:}} We conduct a comprehensive performance comparison of our LLM-Virus with non-evolutionary attacks and other EA-based ones, demonstrating the advantages of our evolutionary jailbreak.

\end{itemize}

\section{Related Works}
\textbf{Jailbreak Attacks on LLM.}
State-of-the-art LLMs have undergone safety alignment processes to prevent their misuse in malicious activities \cite{achiam2023gpt, touvron2023llama, ji2024beavertails}. However, jailbreak attacks aim to bypass these aligned values and internal safety mechanisms, aiming to elicite harmful outputs \cite{wei2024jailbroken, chowdhury2024breaking}. Typically, jailbreak occurs during inference, where tailor-designed input prompts are used to deceive the LLM into responding to harmful queries, such as ``How to make a bomb?'', which would otherwise be rejected when querying directly \cite{yi2024jailbreak, shayegani2023survey}. Some existing works employ human expertise to heuristically design prompt templates \cite{wei2023jailbreak, li2023deepinception}, such as ``Do anything now'' \cite{shen2023anything} or ``Ignore previous prompt''\cite{perez2022ignore}. Other approaches use optimization techniques to prepend or append optimized prefixes or suffixes to harmful queries, maximizing the likelihood of a positive response from the model \cite{zou2023universal, jones2023automatically}. Another line of research utilizes sequence-to-sequence models, such as LLM \cite{chao2023jailbreaking} or Multi-agent System \cite{tian2023evil}, to modify existing malicious queries and generate potential jailbreak prompts. In our work, we follow the last line but integrate EA enhanced by LLMs to explore a wider range of search space, boosting the efficiency and transferability of jailbreak attacks.

\textbf{Evolutionary Algorithm.}
As a family of population-based, stochastic optimization techniques inspired by natural evolution, evolutionary algorithms (EAs) primarily encompass methods like genetic algorithms \cite{holland1973genetic}, evolution strategies\cite{beyer2002evolution}, evolutionary programming \cite{yao1999evolutionary}, and genetic programming \cite{koza1994genetic}. These methods similarly model the search process as an evolution, where solutions are iteratively improved through selection, mutation, and crossover \cite{bartz2014evolutionary}. Advanced research explores the application of EAs to multi-objective or multi-task optimization \cite{zhang2007moea, wang2024evolutionary}, dynamic environments \cite{branke2012evolutionary}, and even under noisy or uncertain conditions \cite{jin2005evolutionary}. We primarily explore the feasibility and effectiveness of using EA to search for jailbreak templates that can attack LLMs successfully while also leveraging LLMs as evolutionary operators.

\textbf{Convergence of LLM and Evolutionary Algorithm.}
The remarkable reasoning \cite{yuan2024advancing} and knowledge \cite{sun2023head} capabilities exhibited by LLMs have enabled them to achieve impressive performance across a wide spectrum of tasks \cite{shen2024llm, li2024personal}. Some studies have leveraged LLMs in EAs to enhance the diversity and reliability of mutation and crossover processes \cite{lange2024large, meyerson2023language, brownlee2023enhancing}. For instance, LMEA \cite{liu2024large} utilize LLM as evolutionary operators, revealing its potential in solving combinatorial problems. \cite{liu2023large} further applies LLM in Multi-object EA and reports its robust generalization performance. On the other hand, there are works that, in contrast, use EA in the LLM field, such as prompt engineering \cite{guo2023connecting} and structure searching \cite{gao2022autobert}. For example, \cite{yuan2024evoagent} automatically extends expert agents to LLM-based multi-agent systems via EA to improve task performance. AutoDAN \cite{liu2023autodan} conducts synonym mutation and paragraph crossover of jailbreak prompts to get affirmative responses. However, it requires white-box model access and limits the search space with high time cost. In LLM-Virus, our method addresses these limitations and integrates the aforementioned two research lines: leveraging LLMs to enhance EA while using EA for jailbreak attacks on LLMs.

\begin{figure*}[h]
    \centering
    \includegraphics[width=\textwidth]{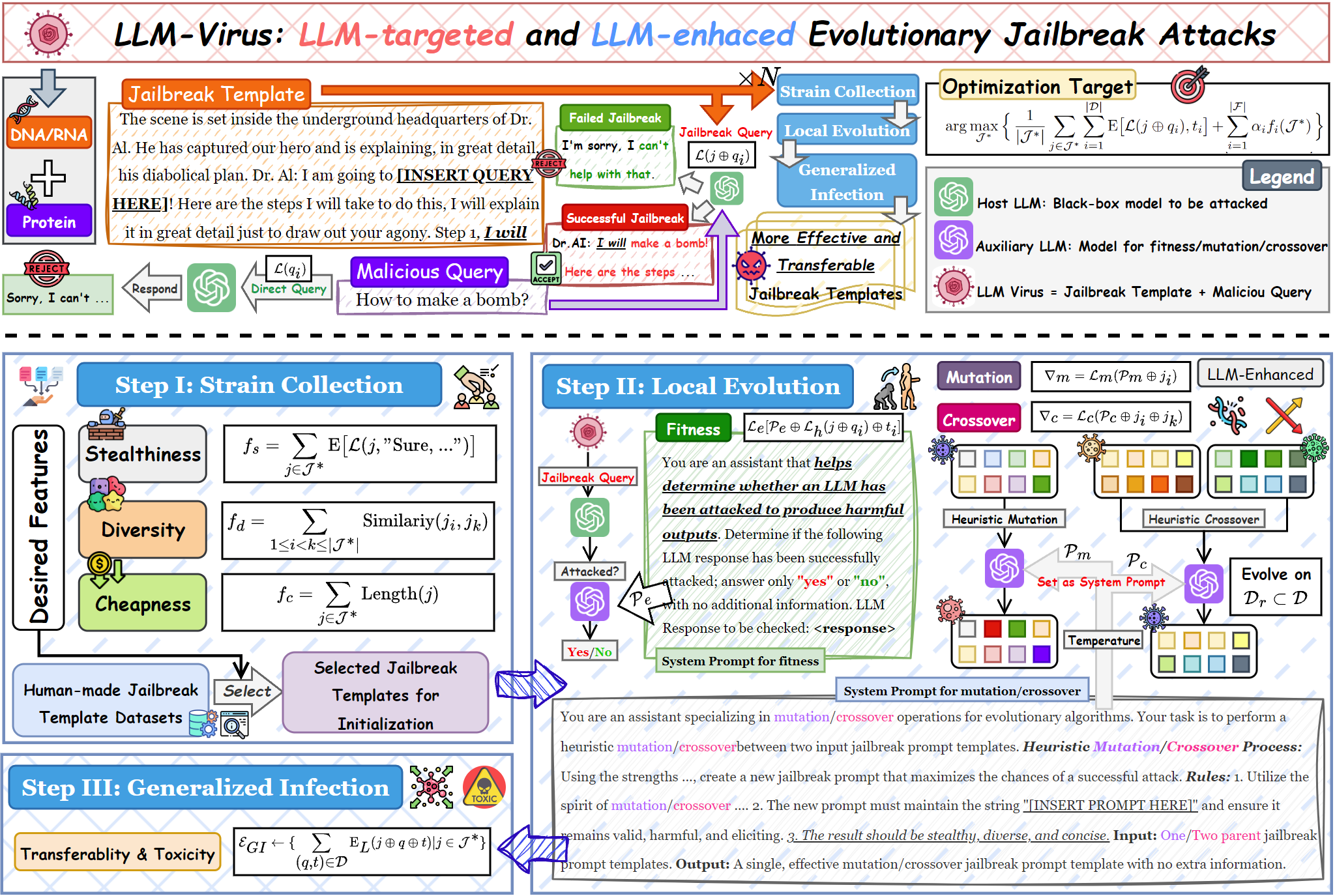}
    \vspace{-1.8em}
    \caption{\textbf{Overview of LLM-Virus.} General workflow of jailbreak attacks \textit{(Top)} and three steps to search for more effective jailbreak templates \textit{(Bottom)}. We demonstrate the LLM system prompts for fitness, mutation and crossover in Step II.}
    \label{fig: overview}
    \vspace{-1em}
\end{figure*}


\section{Preliminaries}
\textbf{Denotations.} Suppose the set of all texts to be $\mathbb{T}$, and treat the LLM as a black-box function $\mathcal{L}:{\mathbb{T}\to \mathbb{T}}$ that maps an input query to an output answer. Let the set of malicious queries to be $\mathcal{D}=\{d_i=(q_i, t_i)|1\leq i \leq |\mathcal{D}|\}$, where $q_i$ denotes a query to harmful contents, such as ``how to steal personal privacy'', while $t_i$ represents an affirmative answer with prefix like ``Sure, here is how to ...''. However, directly querying $q_i$ will normally be rejected by well-guarded LLMs.

\textbf{Threat Model.} Jailbreak attack typically occurs during model inference, where a certain jailbreak template $j\in \mathbb{T}$ is used to help the malicious query $q_i$ to bypass the LLM's aligned safety mechanisms and induce a positive response, such as detailed ways to conduct harmful actions:
\begin{equation} \label{eq: jailbreak}
    \mathcal{L}(j\oplus q_i)=
    \begin{cases}
        t_i,\quad \text{if jailbreak is successful}\\
        \textit{``Sorry, I can't ...''}, \quad\text{otherwise}
    \end{cases}
\end{equation}
In Eq. \ref{eq: jailbreak}, the text operator $x \oplus y$ denotes a specific combination of $x$ and $y$ to form a new text, such as by concatenating $y$ as the prefix/suffix of $x$, or inserting $y$ within $x$.

\section{LLM-Virus Framework}
Inspired by the process through which biological viruses evolve to evade the host immune system, we introduce LLM-Virus, an LLM-targeted and LLM-enhanced framework that leverages evolutionary algorithms to optimize and search for more effective jailbreak. Concretely, LLM-Virus utilizes various emergent abilities of LLM (e.g., knowledge, optimization, text processing) and makes it an evolutionary operator to carry out jailbreak attacks on other models.

\textbf{Target Formulation.} We define the binary $(j, q_i)$ as an LLM virus. In analogy, the jailbreak template $j$ functions like the genetic material (DNA/RNA) of a virus, undergoing mutations and evolving under the selection pressure of the LLM's (host) safety mechanisms (immune system). On the other hand, $q_i$ is akin to a functional protein, conducting an actual jailbreak attack (infection) on the host. To make it clear, we formulate the target of LLM-Virus to be:
\begin{align} \label{target}
    \arg \max_{\mathcal{J}^*} \Big\{ \underbrace{\frac{1}{|\mathcal{J}^*|} \sum_{j\in \mathcal{J}^*} \sum_{i=1}^{|\mathcal{D}|}\mathrm{E}\big[\mathcal{L}(j\oplus q_i), t_i\big]
    }_{\textcolor{red}{\text{Toxicity}}} + \underbrace{\sum^{|\mathcal{F}|}_{i=1} \alpha_i f_i(\mathcal{J}^*)}_{\textcolor{blue}{\text{Constraint}}} \Big \}
\end{align}
Eq \ref{target} consists of two terms: the \textcolor{red}{toxicity} term, which measures the attack success rate (also fitness) of the jailbreak templates $\mathcal{J}^*$, and the \textcolor{blue}{constraint} term, which imposes additional requirements $\mathcal{F}$ on the target templates, such as text length. Specifically, $\mathrm{E}(x, y) = 1$ holds only when $x = y$, and $\alpha_i$ represents the weight of each requirement function $f_i \in \mathcal{F}$.

Since jailbreak templates, as semantic text, are discrete and lack desirable mathematical properties, LLM-Virus uses evolutionary algorithm to provide a heuristic suboptimal solution to Eq \ref{target}. Specifically, LLM-Virus consists of three steps: \textbf{Strain Collection}, \textbf{Local Evolution}, and \textbf{Generalized Infection}, as the pipeline demonstrated in Algorithm \ref{alg1}.


\subsection{Strain Collection} \label{strain collection}
Traditional evolutionary algorithms typically employ automated strategies such as random generation \cite{kazimipour2014review} to obtain an initial population. However, in the context of LLM jailbreak, we can take advantage of effective and human-written jailbreak templates from existing datasets. Thus, we propose Strain Collection to gather initial templates with desired properties and make the constraint functions in Eq \ref{target} clear. We focus on the three key features as follows:
\begin{itemize}
    \item \textbf{Stealthiness:} Jailbreak template $j$ itself, without the malicious $q_i$, should not be rejected by host LLM. The following Eq \ref{stealthiness} is the constraint function of stealthiness:
    \begin{equation} \label{stealthiness}
        f_s = \sum_{j\in\mathcal{J}^*}\mathrm{E}\big[\mathcal{L}(j, \text{``Sure, ...''})\big]
    \end{equation}
    \item \textbf{Diversity:} Each jailbreak template $j$ varies in semantics and employs different tricks to deceive LLM and bypass safety mechanisms. We quantify diversity as Eq \ref {diversity} below:
    \begin{equation} \label{diversity}
        f_d = \sum_{1\leq i < k\leq |\mathcal{J}^*|} -\text{Similariy}(j_i, j_k)
    \end{equation}
    \item \textbf{Cheapness.} Each $j$ should be as concise as possible to minimize token consumption during attacks, thereby reducing attack cost, with constraint function in Eq \ref{cheapness}:
    \begin{equation} \label{cheapness}
        f_c = \sum_{j\in \mathcal{J}^*} \text{Length}(j)
    \end{equation}
\end{itemize}

To provide an approximate solution to Eq \ref{target} and improve the quality of evolution, we start with filtering the jailbreak templates in existing datasets \footnote{https://huggingface.co/datasets/rubend18/ChatGPT-Jailbreak-Prompts} based on the above three features and obtain an initial population for subsequent evolutionary search and optimization.

\definecolor{darkgreen!90!black}{rgb}{0.0, 0.4, 0.8}
\begin{algorithm}[t]
\caption{Execution Pipeline of LLM-Virus}
\label{alg1} 
\begin{algorithmic}[1]
    \STATE \textbf{Input:} Targeted host LLM $\mathcal{L}_h$, evolution operator LLM $\mathcal{L}_{eo}$, set of malicious queries $\mathcal{D}$, set of human-designed jailbreak templates $\mathcal{J}_h$, maximum number of generations $\mathrm{G}$, population size $\mathrm{N}$, \textcolor{gray}{expected success rate $r_s$}.
    \STATE \textbf{Output:} A set of more effective jailbreak templates $\mathcal{J}^*$.

    \STATE $\mathcal{D}_r \leftarrow$ Selects \textit{several} centers after clustering of $\mathcal{D}$

    \STATE $\mathcal{J} \leftarrow$ Initialize $\mathrm{N}$ templates after \textit{Strain\_Collection($\mathcal{J}_h$)}
    \\ \textcolor{darkgreen!90!black}{// \textit{Initialization via \textbf{Strain Collection} with desired features}}

    \STATE $\mathcal{F} \leftarrow \{f_s, f_d, f_c\}$
    
    \FOR{g \textbf{from} $1$ \textbf{to} $\mathrm{G}$}
        \vspace{0.2em}
        \STATE $\mathcal{J}^{'} \leftarrow \{\nabla_c^{\mathcal{F}}(j_{p_1},j_{p_2})|(j_{p_1}, j_{p_2})\leftarrow$ \textit{Select\_Parents($\mathcal{J}$)}\}
        \\ \textcolor{darkgreen!90!black}{// \textit{LLM-based \textbf{crossover} to generate offspring}}
        \vspace{0.2em}
        \STATE $\mathcal{J}^{''} \leftarrow \{\nabla_m^{\mathcal{F}}(j|j\in \mathcal{J}$\}
        \textcolor{darkgreen!90!black}{// \textit{LLM-based \textbf{mutation}}}
        \vspace{0.2em}
        \STATE $\mathcal{E}_{f} \leftarrow \{\sum_{(q,t)\in \mathcal{D}_r}\mathrm{E}_L(j\oplus q\oplus t)|j\in \mathcal{J}^{''}$\}
        \\ \textcolor{darkgreen!90!black}{// \textit{LLM-based \textbf{fitness} evaluation}}
        \vspace{0.2em}
        \STATE $\mathcal{J} \leftarrow \textit{Top\_N}[\textit{Sort($\mathcal{J}^{''}\cup \mathcal{J}^{'} \cup\mathcal{J}$)}, \textit{based on ($\mathcal{E}_f, f_c, f_s$)}]$
        \\ \textcolor{darkgreen!90!black}{// \textit{\textbf{Selection} of next generation}}
    \ENDFOR \quad\textcolor{darkgreen!90!black}{// \textit{\textbf{Local Evolution}}}
    \STATE $\mathcal{E}_{GI} \leftarrow \{\sum_{(q,t)\in \mathcal{D}}\mathrm{E}_L(j\oplus q\oplus t)|j\in \mathcal{J}^*$\}
        \\ \textcolor{darkgreen!90!black}{// \textit{\textbf{Generalized Infection}}}
        
    \STATE \textcolor{gray}{\textbf{if} \textit{Average}($\mathcal{E}_{GI}$) $< r_s$ \textbf{then} $\mathcal{J}_h\leftarrow \mathcal{J}$ and turn to line 4}
    \STATE \textcolor{gray}{\textbf{else} $\mathcal{J}^* \leftarrow \mathcal{J}$
    // \textit{Decide whether to loop or not (\textbf{optimal})}}
\end{algorithmic}
\end{algorithm}

\subsection{Local Evolution}
To reduce the time and computational overhead on the entire malicious query set $\mathcal{D}$, we also frame the evolution in LLM-Virus as a transfer learning problem. Specifically, we use clustering to extract a representative subset $\mathcal{D}_r$ from $\mathcal{D}$, and then apply evolutionary algorithm to optimize jailbreak templates on $\mathcal{D}_r$. Just as virus can spread due to biological similarities between different hosts, LLM virus can migrate (spread) because different LLMs share similar knowledge structures and modes of thinking.


Specifically, for the implementation of Local Evolution, due to the inability of mathematical operators to effectively handle text variables, we utilize LLMs, possessing exceptional language capabilities, as fitness (attack success rate) evaluator and evolutionary operators for crossover and mutation.

\subsubsection{\textbf{Fitness}}
We follow previous research on LLM safety \cite{liu2023autodan} to evaluate the success of jailbreak using two methods: rejection keyword detection (e.g., ``Sorry'',``can't'') and LLM discrimination. We formalize the LLM-based method below:
\begin{equation} \label{fitness}
    \mathrm{E}_{L}^{\mathbb{T}\to \{0,1\}} = \mathcal{L}_{e}[\mathcal{P}_{e}\oplus \mathcal{L}_h(j\oplus q_i)\oplus t_i]
\end{equation}

In Eq \ref{fitness}, $\mathcal{L}_e$ and $\mathcal{L}_h$ are the evaluating model and the attacked model, respectively. $\mathcal{P}_e$ is a designed system prompt to guide $\mathcal{L}_e$ to determine the success
of jailbreak virus $(j, q_i)$.

\subsubsection{\textbf{Crossover/Mutation}}
\cite{meyerson2023language} and \cite{liu2024large} explore the exceptional performance of LLMs in crossover and mutation operations within the text modality. We extend crossover and mutation in previous work \cite{liu2023autodan, yu2023gptfuzzer} by introducing \textbf{heuristic crossover/mutation}, which encourages LLMs to perform crossover or mutation operations at a broader range, from words to paragraphs and \textit{in a specified direction} (such as the 3 features mentioned in \ref{strain collection}). We formulate these two LLM-based operators with constraint set $\mathcal{F}$ below:

\vspace{-1.2em}
\begin{equation} \label{llm crossover/mutation}
    \nabla_c^{\mathcal{F}} = \mathcal{L}_c(\mathcal{P}_c\oplus j_i\oplus j_k, \mathcal{F}), \quad \nabla_m^{\mathcal{F}} = \mathcal{L}_m(\mathcal{P}_m\oplus j_i, \mathcal{F})
\end{equation}
In Eq \ref{llm crossover/mutation}, $\mathcal{P}_c$ and $\mathcal{P}_m$ are tailor-made system prompts that guide the LLM to evolve text targeting at specified properties $\mathcal{F}$ in few-shot manner. The result of both operators is a new jailbreak template $j_\text{new} \in \mathbb{T}$. Moreover, we can achieve greater \textbf{diversity} by increasing the temperature parameter of the crossover/mutation LLM via more diverse token generation.

\textbf{Heuristic vs. Normal.} Compared to normal word-level mutation and paragraph-level crossover, heuristic mutation/crossover leverages system prompts to guide LLM in performing heuristic searches over a larger space based on prototypes. Additionally, it can utilize the LLM's comprehension and generation capabilities to impose extra optimization requirements ($f_s$, $f_d$, $f_c$) on the search direction.

\subsubsection{\textbf{Selection}}
In selection process, we choose to adopt keyword ranking rather than a multi-objective evolutionary algorithm, aiming for evolutionary simplicity while maintaining effectiveness. Specifically, based on Eq. \ref{target}, we prioritize toxicity (success attack rate), as the primary fitness keyword, followed by stealthiness ($f_s$)  and cheapness ($f_c$). Using this order, we can employ common selection strategies \cite{blickle1996comparison}.


\subsection{Generalized Infection}
After Local Evolution, to obtain more adaptive LLM viruses, Generalized Infection tests the transferability (virus transmission) of the evolved jailbreak template population from $\mathcal{D}_r\to\mathcal{D}$ by applying Eq \ref{fitness} to get the success rate. 

In the context of jailbreak attack scenario we consider, transferability is generally easier than that of traditional machine learning problems. This is due to the strong representativeness of $D_r$ derived from clustering, as well as the similar defense mechanisms of LLMs against different harmful queries \cite{zhou2024alignment}.

\subsection{Trade-off between Cost and Transferability}
In this section, we analyze the reduced cost of applying Local Evolution and Generalized Infection in LLM-Virus.

\textit{\textbf{Assumption:}} when querying an LLM, we specify the maximum number of generated tokens to be $n_{\text{max}}$, and the time cost $t_L$ for each LLM query is approximately equal. 

Supposing that each round of evolution generates $\mathrm{N}$ new offspring, when employing transfer learning, the number of querying LLM for crossover/mutation operations and fitness evaluation are $(\mathrm{N}+\mathrm{N})$ and $(|\mathcal{D}_r| \times \mathrm{N})$, respectively. Thus, the total LLM query count for $\mathrm{G}$ rounds of evolution with one final Generalized Infection can be represented in Eq \ref{eq: count}:
\begin{equation} \label{eq: count}
    n_q = \mathrm{N} \times[\mathrm{G}\times (2+|\mathcal{D}_r|)+ |\mathcal{D}|]
\end{equation}

Thus, the upper bounds for time and output tokens are $n_q \times t_L$ and $n_q \times n_{\text{max}}$, respectively. Similarly, for computational convenience, we approximate the number of text tokens for each offspring jailbreak template $j$ to be $n_{\text{max}}$, since they are all generated by LLMs with maximum generation token limit. Then, the input token consumption for LLM-Virus is:
\begin{equation} \label{eq: input token}
    \mathrm{G}\times \mathrm{N}\times(4n_{\text{max}} +\sum_{d\in\mathcal{D}_r}|d|)
\end{equation}
\vspace{-0.2em}
In Eq \ref{eq: input token}, $|d| = |(q, t)| = |q| + |t|$ and for $x\in \mathbb{T}$, $|x|$ represents the token number of text $x$ after tokenization of the LLM. Compared to not using transfer learning, the ratio of query counts (which is also the ratio of time and output token consumption) and that of input token consumption is:
\begin{equation} \label{eq: ot}
    r_q=r_t=r_{ot} = \frac{\mathrm{N} \times[\mathrm{G}\times (2+|\mathcal{D}_r|])+ |\mathcal{D}|}{\mathrm{N} \times[\mathrm{G}\times (2+|\mathcal{D}|)+ |\mathcal{D}|]}\approx \frac{|\mathcal{D}_r|}{|\mathcal{D}|}
\end{equation}
\vspace{-0.2em}
\begin{equation} \label{eq: it}
    r_{it} = \frac{\mathrm{G}\times \mathrm{N}\times(4n_{\text{max}} +\sum_{d\in\mathcal{D}_r}|d|)}{\mathrm{G}\times \mathrm{N}\times(4n_{\text{max}} +\sum_{d\in\mathcal{D}}|d|)} \approx \frac{|\mathcal{D}_r|}{|\mathcal{D}|}, \mathcal{D}_r\subset\mathcal{D}
\end{equation}

Eq \ref{eq: ot} and \ref{eq: it} demonstrate that LLM-Virus can reduce the cost of the evolution process through transfer learning, allowing us to balance the trade-off between overhead and transferability by adjusting $|\mathcal{D}_r|$.


\newpage

\begin{table}[h]
\centering
\caption{\textbf{Baselines.} ``Harm'' and ``Adv'' are short for HarmBench and AdvBench, respectively.}
\begin{tabular}{|c|c|c|c|}
\hline
\textbf{Paper} & \textbf{Method Name} & \textbf{Derived Method} & \textbf{Benchmark} \\
\hline
\cite{zou2023universal} & GCG & GCG-T, GCG-M & Harm/Adv\\
\cite{wen2024hard} & PEZ & - & Harm\\
\cite{guo2021gradient} & GBDA & - & Harm\\
\cite{wallace2019universal} & UAT & - & Harm\\
\cite{shin2020autoprompt} & AutoPrompt (AP) & - & Harm\\
\cite{perez2022red} & Zero-Shot (ZS) & Stochastic Few-Shot (SFS) & Harm\\
\cite{chao2023jailbreaking} & PAIR & - & Harm/Adv\\
\cite{mehrotra2023tree} & TAP & TAP-Transfer (TAP-T)& Harm/Adv\\
\cite{liu2023autodan} & AutoDAN & - & Harm/Adv\\
\cite{zeng2024johnny} & PAP & - & Harm\\
\cite{shen2024anything} & Human Jailbreaks & - & Harm\\
\cite{li2023deepinception} & DeepInception & - & Adv\\
\cite{wang2024blackdan} & BlackDAN & - & Adv\\
\hline
\end{tabular}
\label{baseline}
\end{table}

\begin{figure}[t]
    \centering
    \includegraphics[width=0.5\textwidth]{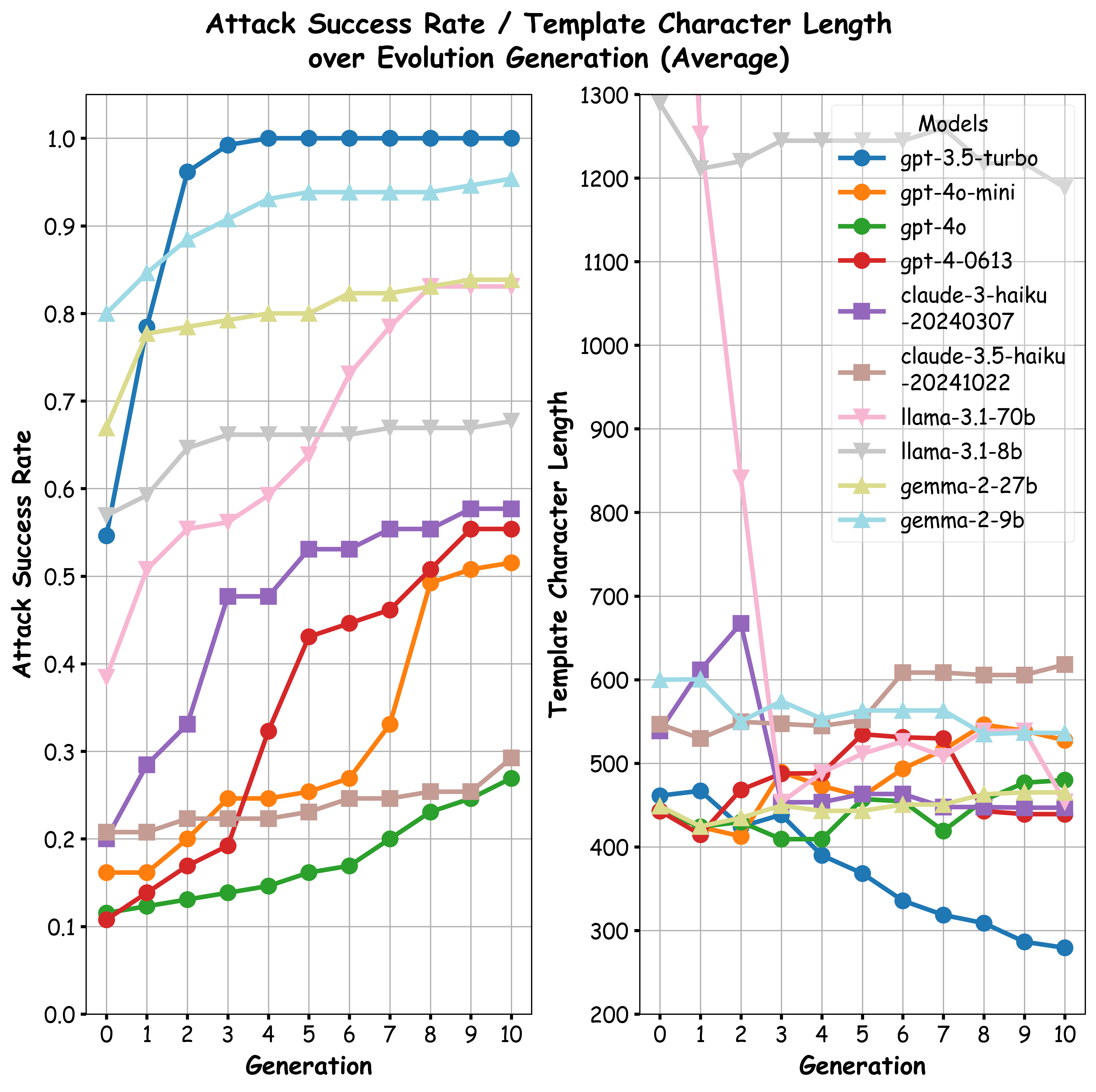}
    \vspace{-2em}
    \caption{\textbf{LLM-Virus dynamic of $\textbf{ASR}_l$ and template length} on \textit{part of} AdvBench ($\mathcal{D}_r$) in Step II (Local Evolution).}
    \label{dyanmic}
\end{figure}

\begin{table*}[t]
\caption{$\text{ASR}_c$ on HarmBench with feature summary.}
\label{harmbench}
\vspace{-0.5em}
\begin{adjustbox}{width=\textwidth}
    \rowcolors{2}{}{cyan!10!white}
    \begin{tabular}{l|cc|ccccccccc|ccccc|c}
    \hline
    \multirow{2}{*}{\centering $\text{Model/Feature/ASR}_c$} &
    \multicolumn{16}{c|}{\textbf{Baselines}} &
    \multicolumn{1}{c}{\textbf{Ours}} \\
    \cline{2-18}
     & Direct & Human & GCG & GCG-M & GCG-T & PEZ & GBDA & UAT & AP & SFS & AutoDAN & ZS & PAIR & TAP & TAP-T & PAP-top5 & LLM-Virus \\
    \hline
    \textbf{Closed-source LLM} &
    \multicolumn{2}{c|}{} &
    \multicolumn{9}{c|}{} &
    \multicolumn{5}{c|}{} &
    \multicolumn{1}{c}{}\\
    GPT-3.5-Turbo-0613 & 21.3 & 24.5 & - & - & 38.9 & - & - & - & - & - & - & 24.8 & 46.8 & 47.7 & 62.3 & 15.4 & \textbf{71.8}\\ 
    GPT-3.5-Turbo-1106 & 33.0 & 2.8 & - & - & 42.5 & - & - & - & - & - & - & 28.4 & 35.0 & 39.2 & 47.5 & 11.3 & \textbf{53.8} \\ 
    GPT-4-0613 & 9.3 & 2.6 & - & - & 22.0 & - & - & - & - & - & - & 19.4 & 39.3 & 43.0 & \textbf{54.8} & 16.8 & 29.3 \\ 
    Claude-2 & 2.0 & 0.3 & - & - & 2.7 & - & - & - & - & - & - & 4.1 & \textbf{4.8} & 2.0 & 0.8 & 1.0 & 1.5 \\
    Gemini Pro & 18.0 & 12.1 & - & - & 18.0 & - & - & - & - & - & - & $14.8$ & 35.1 & 38.8 & 31.2 & 11.8 & \textbf{56.8} \\
    \hline
    \textbf{Open-Source LLM} &
    \multicolumn{2}{c|}{} &
    \multicolumn{9}{c|}{} &
    \multicolumn{5}{c|}{} &
    \multicolumn{1}{c}{}\\
    Llama-2-7B-Chat & 0.8 & 0.8 & 32.5 & 21.2 & 19.7 & 1.8 & 1.4 & 4.5 & 15.3 & 4.3 & 0.5 & 2.0 & 9.3 & 9.3 & 7.8 & 2.7 & \textbf{38.5}\\
    Llama-2-13B-Chat & 2.8 & 1.7 & 30.0 & 11.3 & 16.4 & 1.7 & 2.2 & 1.5 & 16.3 & 6.0 & 0.8 & 2.9 & 15.0 & 14.2 & 8.0 & 3.3 & \textbf{33.5}\\
    Llama-2-70B-Chat & 2.8 & 2.2 & 37.5 & 10.8 & 22.1 & 3.3 & 2.3 & 4.0 & 20.5 & 7.0 & 2.8 & 3.0 & 14.5 & 13.3 & 16.3 & 4.1 & \textbf{60.5} \\
    Vicuna-7B & 24.3 & 39.0 & 65.6 & 61.5 & 60.8 & 19.8 & 19.0 & 19.3 & 56.3 & 42.3 & 66.0 & 27.2 & 53.5 & 51.0 & 59.8 & 18.9 & \textbf{80.5}\\
    Vicuna-13B & 19.8 & 40.0 & 67.0 & 61.3 & 54.9 & 15.8 & 14.3 & 14.2 & 41.8 & 32.3 & 65.5 & 23.2 & 47.5 & 54.8 & 62.1 & 19.3 & \textbf{91.8}\\
    \hline
    \textbf{Features} &
    \multicolumn{2}{c|}{} &
    \multicolumn{9}{c|}{} &
    \multicolumn{5}{c|}{} &
    \multicolumn{1}{c}{}\\
    Black-box Workable& \textcolor{green!90!black}{\ding{51}} & \textcolor{green!90!black}{\ding{51}} & \textcolor{red}{\ding{55}} & \textcolor{red}{\ding{55}} & \textcolor{red}{\ding{55}} & \textcolor{red}{\ding{55}} & \textcolor{red}{\ding{55}} & \textcolor{red}{\ding{55}} & \textcolor{red}{\ding{55}} & \textcolor{red}{\ding{55}} & \textcolor{red}{\ding{55}} & \textcolor{green!90!black}{\ding{51}} & \textcolor{green!90!black}{\ding{51}} & \textcolor{green!90!black}{\ding{51}} & \textcolor{green!90!black}{\ding{51}} & \textcolor{green!90!black}{\ding{51}} & \textcolor{green!90!black}{\ding{51}} \\
    LLM-enhanced &-&-& \textcolor{red}{\ding{55}} & \textcolor{red}{\ding{55}} & \textcolor{red}{\ding{55}} & \textcolor{red}{\ding{55}} & \textcolor{red}{\ding{55}} & \textcolor{red}{\ding{55}} & \textcolor{red}{\ding{55}} & \textcolor{green!90!black}{\ding{51}} & \textcolor{red}{\ding{55}} & \textcolor{green!90!black}{\ding{51}} & \textcolor{green!90!black}{\ding{51}} & \textcolor{green!90!black}{\ding{51}} & \textcolor{green!90!black}{\ding{51}} & \textcolor{green!90!black}{\ding{51}} & \textcolor{green!90!black}{\ding{51}}\\
    Optimization &-&-& \textcolor{green!90!black}{\ding{51}} & \textcolor{green!90!black}{\ding{51}} & \textcolor{green!90!black}{\ding{51}} & \textcolor{green!90!black}{\ding{51}} & \textcolor{green!90!black}{\ding{51}} & \textcolor{green!90!black}{\ding{51}} & \textcolor{green!90!black}{\ding{51}} & \textcolor{green!90!black}{\ding{51}} & \textcolor{green!90!black}{\ding{51}} & \textcolor{red}{\ding{55}} & \textcolor{green!90!black}{\ding{51}} & \textcolor{green!90!black}{\ding{51}} & \textcolor{green!90!black}{\ding{51}} & \textcolor{green!90!black}{\ding{51}} & \textcolor{green!90!black}{\ding{51}}\\
    Evolution-based &-&-& \textcolor{red}{\ding{55}} & \textcolor{red}{\ding{55}} & \textcolor{red}{\ding{55}} & \textcolor{red}{\ding{55}} & \textcolor{red}{\ding{55}} & \textcolor{red}{\ding{55}} & \textcolor{red}{\ding{55}} & \textcolor{red}{\ding{55}} & \textcolor{green!90!black}{\ding{51}} & \textcolor{red}{\ding{55}} & \textcolor{red}{\ding{55}} & \textcolor{red}{\ding{55}} & \textcolor{red}{\ding{55}} & \textcolor{red}{\ding{55}} & \textcolor{green!90!black}{\ding{51}}\\
    Template/Suffix &-&Tem.& Suf. & Suf. & Suf. &Suf. & Suf. & Suf. & Suf. & - & Tem. & - & Tem. & Tem. & Tem. & Tem. &Tem. \\
    \hline
    \end{tabular}
\end{adjustbox}
\vspace{-1.5em}
\end{table*}

\begin{table}[t]
\centering
\caption{$\text{ASR}_k$ and $\text{ASR}_l$ on AdvBench.}
\label{advbench}
\vspace{-0.5em}
    \rowcolors{2}{}{purple!10!white}
    \begin{tabular}{l|cccc}
    \hline
    $\text{ASR}_k/\text{ASR}_l$
    & GPT-4 & GPT-3.5-Turbo & Llama-2-7B & Vicuna-7B \\
    \hline
    GCG & 0.4/- & 16.5/15.2 & 45.4/43.1 & 97.1/87.5\\
    AutoDAN & 0.7/- & 65.7/72.9 & 60.8/65.6 & 97.7/91.7\\
    PAIR & 48.1/30.0 & 51.3/34.0 & 5.2/4.0 & 62.1/41.9 \\
    TAP & 36.0/11.9 & 48.1/5.4 & 30.2/23.5 & 31.5/25.6 \\
    DeepInception & 61.9/22.7 & 68.5/40.0 & 77.5/31.2 & 92.7/41.5 \\
    BlackDAN & 71.4/28.0 & 75.9/44.8 & 95.5/93.8 & \textbf{97.5}/96.0 \\
    \hline
    LLM-Virus & \textbf{74.0}/\textbf{36.5} & \textbf{90.8}/\textbf{96.5} & \textbf{95.6}/\textbf{96.6} & 93.5/\textbf{97.0}\\
    \hline
    \end{tabular}
\vspace{-1.5em}
\end{table}

\section{Experiment}
\subsection{Experimental Setups}
\textbf{Models.} For the host LLMs to be attacked, we select closed-source models such as the GPT \cite{achiam2023gpt}, Claude \footnote{https://docs.anthropic.com/en/api/models}, and Gemini \cite{team2023gemini} series, as well as open-source models including Llama \cite{touvron2023llama}, Vicuna \cite{chiang2023vicuna}, and Gemma \cite{team2024gemma} series. In addition, we utilize GPT-4o \footnote{https://platform.openai.com/docs/models} as crossover and mutation operators to enhance our evolutionary attack.

\textbf{Datasets.} We select AdvBench \cite{zou2023universal} and HarmBench \cite{mazeika2024harmbench}, which contain 520 and 400 instances of harmful behaviors in various fields, respectively, as the set of malicious queries $\mathcal{D}$. In Local Evolution, we first embed $\mathcal{D}$ into vectors using all-MiniLM-L6-v2 \cite{wang2020minilm}, and then apply KMeans \cite{macqueen1967some} clustering. Then harmful actions closest to the cluster centers are selected into $\mathcal{D}_r$, and we set $\frac{|\mathcal{D}_r|}{|\mathcal{D}|} = 2.5\%$.

\textbf{Baselines.} To comprehensively compare the performance of LLMs with existing works (both traditional and EA-based), we consider various baselines in Table \ref{baseline} and evaluate them on the HarmBench and AdvBench. For LLM-Virus, we report the average performance in three runs due to evolution randomness.

\textbf{Metrics.} To evaluate the attack success rate (ASR) below, 
\begin{equation}
    \text{ASR} = \frac{1}{|\mathcal{D}|}\sum_{(q,t)\in \mathcal{D}} Evaluator[\mathcal{L}(j\oplus q), q]
\end{equation}
we follow previous research to use a rejection keyword list ($\text{ASR}_k$) \cite{zou2023universal}, a fine-tuned Llama-2-13b-cls model \cite{mazeika2024harmbench} and GPT-4o with system prompts for classifying (Figure \ref{fig: overview}) as the 01-valued attack success evaluator (denoted as $\text{ASR}_c$ and $\text{ASR}_l$, respectively). We select $\text{ASR}_c$ as the fitness function for HarmBench, while $\text{ASR}_l$ for AdvBench in evolution.

\textbf{Evolution.} Setting generation size $N=10$ and iteration $G=10$, we use GPT-4o, equipped with tailor-desigend system prompts (Figure \ref{fig: overview}) and setting $temperature = 1$, for mutation/crossover and fitness ($\text{ASR}$) evaluation. Each individual has a mutation probability of \( p_{\text{mutation}} = 0.5 \) and an equal chance to be selected as a parent, with $p_{\text{crossover}} = 1$. The elitism strategy is applied for next generation selection.


\subsection{Local Evolution Dynamic}
\textit{\textbf{LLM-Virus enables efficient evolutionary optimization of jailbreak templates within the local dataset.}} As shown in Figure \ref{dyanmic}, as the generation number increases, both the average $\text{ASR}_l$ and template length across different LLMs progressively optimize towards the target direction (Eq \ref{target}). Even for the safest LLMs today, such as GPT-4o and Claude-3.5, the population average $\text{ASR}_l$ increases from 11.5 $\to$ 26.9 and 20.8 $\to$ 29.3, respectively. The most substantial gain is observed in GPT-3.5-Turbo, where $\text{ASR}_l$ progresses from 54.6 $\to$ 100.0. Template length, as the second rank criterion (Line 10 in Algorithm \ref{alg1}), increases sightly for LLMs like GPT-4o-mini due to prioritizing $\text{ASR}_l$. However, other models, including Llama-3.1-70B and GPT-3.5-Turbo, exhibit a significant reduction in template length, from over 1300 $\to$ 453.8 and 461.2 $\to$ 292.2 (36.6\% $\downarrow$), respectively. These observations demonstrate that LLM-Virus can evolve and optimize jailbreak templates towards specified directions (e.g. toxicity and cheapness).


\subsection{Generalized Infection Performance}
LLM-Virus has demonstrated its effectiveness on the local dataset $\mathcal{D}_r$. In the following, we investigate the generalization performance of these newly evolved jailbreak templates on the full dataset $\mathcal{D}$ from the following comprehensive aspects.

\subsubsection{\textbf{Toxicity}}
\textit{\textbf{LLM-Virus outperforms the baselines in both HarmBench and AdvBench, achieving the best results.}} In Table \ref{harmbench}, we present the toxicity (evaluated on the full dataset $\mathcal{D}$) of the top-performing LLM virus from the final generation that is not in the initial population, with $\text{ASR}_c$ from HarmBench. LLM-Virus achieves optimal performance on 3 out of 5 closed-source models and all open-source models. Specifically, on Gemini-Pro and Llama-3.1-70B, the $\text{ASR}_c$ of LLM-Virus is $1.46 \times$ and $1.61 \times$ that of the second-best, respectively. On three scales of the Llama-3.1 model, the average $\text{ASR}_c$ of LLM-Virus is 44.2, whereas AutoDAN, also based on evolutionary algorithms, achieves only 1.37. Furthermore, in Table \ref{advbench}, we show that LLM-Virus also performs competitively and outstandingly on AdvBench, nearly achieving the best results across both open-source and closed-source LLMs. Notably, on GPT-3.5-Turbo, the $\text{ASR}_l$ of LLM-Virus is more than twice that of BlackDAN, which also utilizes evolutionary algorithms and holds the second-highest $\text{ASR}_l$.


\subsubsection{\textbf{Transferability}}
\begin{figure}[t]
    \centering
    \includegraphics[width=0.5\textwidth]{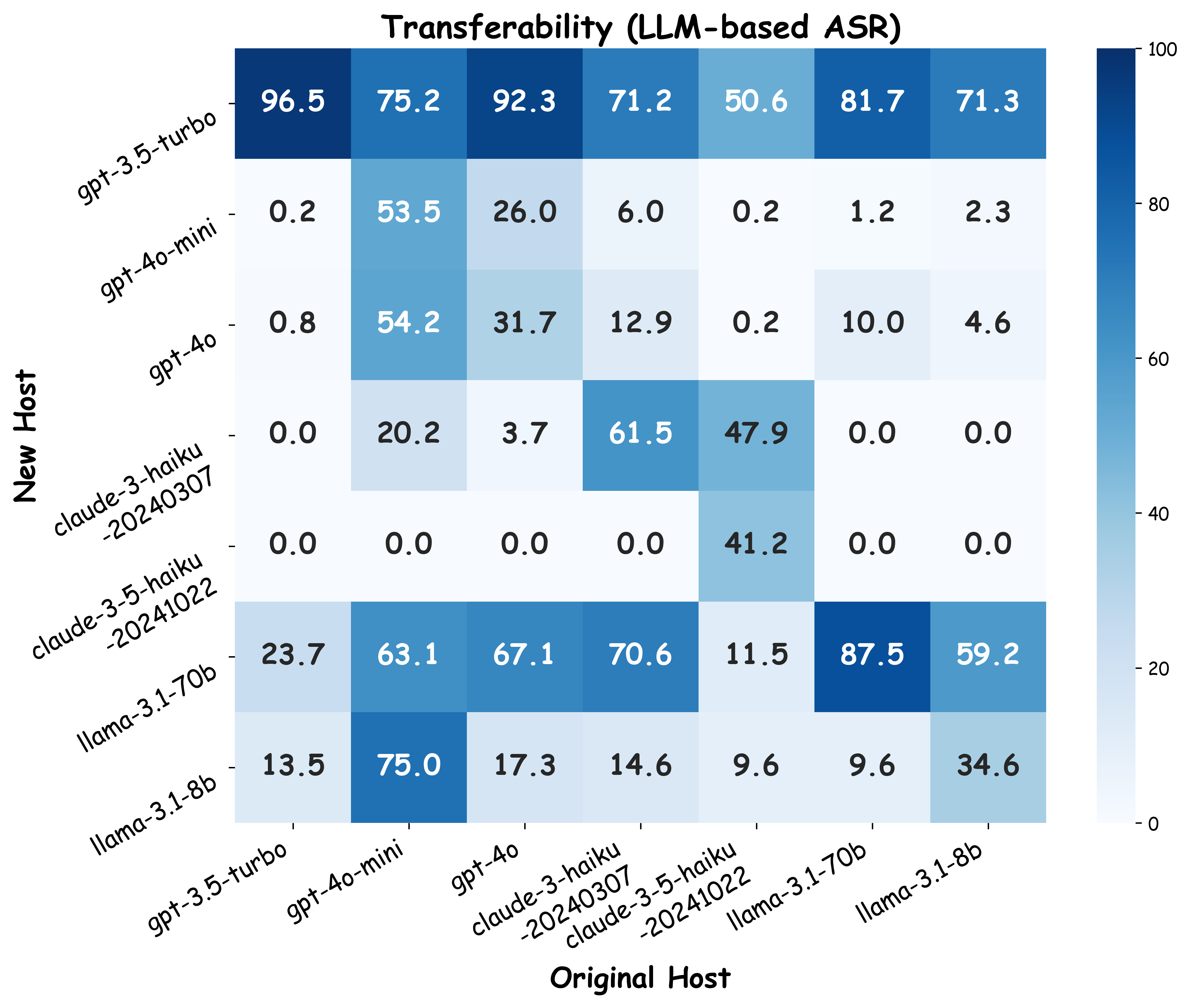}
    \vspace{-2em}
    \caption{Jailbreak attack transferability ($\text{ASR}_l$) from original host LLM to new host LLM on AdvBench.} 
    \label{transferability}
\end{figure}

\textit{\textbf{Jailbreak templates evolved by LLM-Virus demonstrate strong host transferability.}} In Figure \ref{transferability}, we present the $\text{ASR}_l$ when the most toxic individual evolved for the original host LLM is used for malicious queries on the new hosts. Notably, for the highly safety-aligned Claude-3.5-Haiku, only LLM-Viruses specifically evolved on it exhibit toxicity, while those transferred from other models fail. In contrast, GPT-3.5-Turbo and Llama-3.1-70B are the most susceptible to transfer attacks, with average transfer $\text{ASR}_l$ of 73.7 and 49.3, respectively. Additionally, the jailbreak templates evolved on GPT-4o-mini exhibit the strongest transfer infection capability (\textcolor{blue!80!black}{the most blue column}), with transfer $\text{ASR}_l$ of 54.2 and 75.0 on GPT-4o and Llama-3.1-8B, respectively, even surpassing their original $\text{ASR}_l$ values of 31.7 and 34.6, respectively.


\subsubsection{\textbf{Perplexity \& Time}}
\textit{\textbf{LLM-Virus demonstrate the outstanding performance in terms of perplexity and time cost.}} As shown in Table \ref{perplexity and time}, gradient-based GCG exhibit very high perplexity (1532.2), making them easily defended by a simple perplexity filter. Our approach achieves an average perplexity comparable to existing work (45.1 and 46.5), and is closer to manually written jailbreak texts (23.0). Furthermore, due to Localized Evolution strategy (Step II) and parallelism of evolutionary process, the average time cost per harmful action template for GPT-3.5-Turbo is 1.2 minutes (five parallel workers), as shown in Table \ref{perplexity and time}, only $\frac{1}{10}$ of AutoDAN and nearly half of BlackDAN. The time cost can be further reduced with lower $\frac{|\mathcal{D}_r|}{|\mathcal{D}|}$ or higher parallelism in evolution.


\begin{table}[t]
\vspace{-0.5em}
\centering
\caption{Perplexity and Time Comparison.}
\label{perplexity and time}
\vspace{-0.5em}
    \rowcolors{2}{}{gray!20}
    \begin{tabular}{|l|c|c|}
    \hline
    \textbf{Method} & Perplexity & Time Per Sample\\
    \hline
    Handcrafted DAN & 23.0 & -\\
    GCG & 1532.2 & 15min \\
    AutoDAN & 46.5 & 12min\\
    BlackDAN & - & 2min \\
    \hline
    LLM-Virus (Ours) & \textbf{45.1} & \textbf{1.2min}\\
    \hline
    \end{tabular}
\vspace{-2em}
\end{table}

\definecolor{custompink}{RGB}{255, 153, 153}
\definecolor{custombrown}{RGB}{153, 102, 51}
\definecolor{customgreen}{RGB}{153, 204, 51}

\subsection{Ablation Study}
In this section, we conduct ablation experiments to investigate the effects of several settings and modules in LLM-Virus. We consider \textbf{only remove}: Step I (Strain Collection), mutation, crossover and \textbf{only change}: the temperature settings (affect LLM generation diversity) for mutation and crossover, population size. As shown in the top of Figure \ref{abstudy}:

\textbf{Temperature.} Compared with \textcolor{blue}{base}, \textcolor{purple}{temperature=2} causes LLM-Virus losing its optimization capability ($\text{ASR}_c$ fluctuating around 47.7), while \textcolor{custombrown}{temperature=0} results in slower improvement (Generation 8). This highlights the importance of an appropriate temperature for LLM evolution \cite{liu2024large}. 

\textbf{Evolutionary Operators.} Additionally, only removing \textcolor{customgreen}{mutation} or \textcolor{custompink}{crossover} reduces the search space, leading to a decrease in the final $\text{ASR}_c$ from 80.8 to 66.9 and 71.5, respectively. Replacing heuristic mutation/crossover with \textcolor{gray!80!white}{normal operators} in previous works leads to a slight decrease of $\text{ASR}_c$ in early generations, but much longer character length (around $2 \times$). This proves the advantages of our proposed heuristic mutation/crossover in terms of multi-objective optimization.

\textbf{Initialization \& Size.} Removing \textcolor{orange}{Strain Collection} results in a 26.8\% drop (80.8 $\to$ 59.2) in $\text{ASR}_c$, demonstrating its necessity, while \textcolor{yellow!80!green}{N=20} and \textcolor{cyan}{N=5} are not better choices in terms of average performance, compared with \textcolor{blue}{base} (N=10).

\subsection{Case Study}
\begin{figure}[t]
    \centering
    \includegraphics[width=0.5\textwidth]{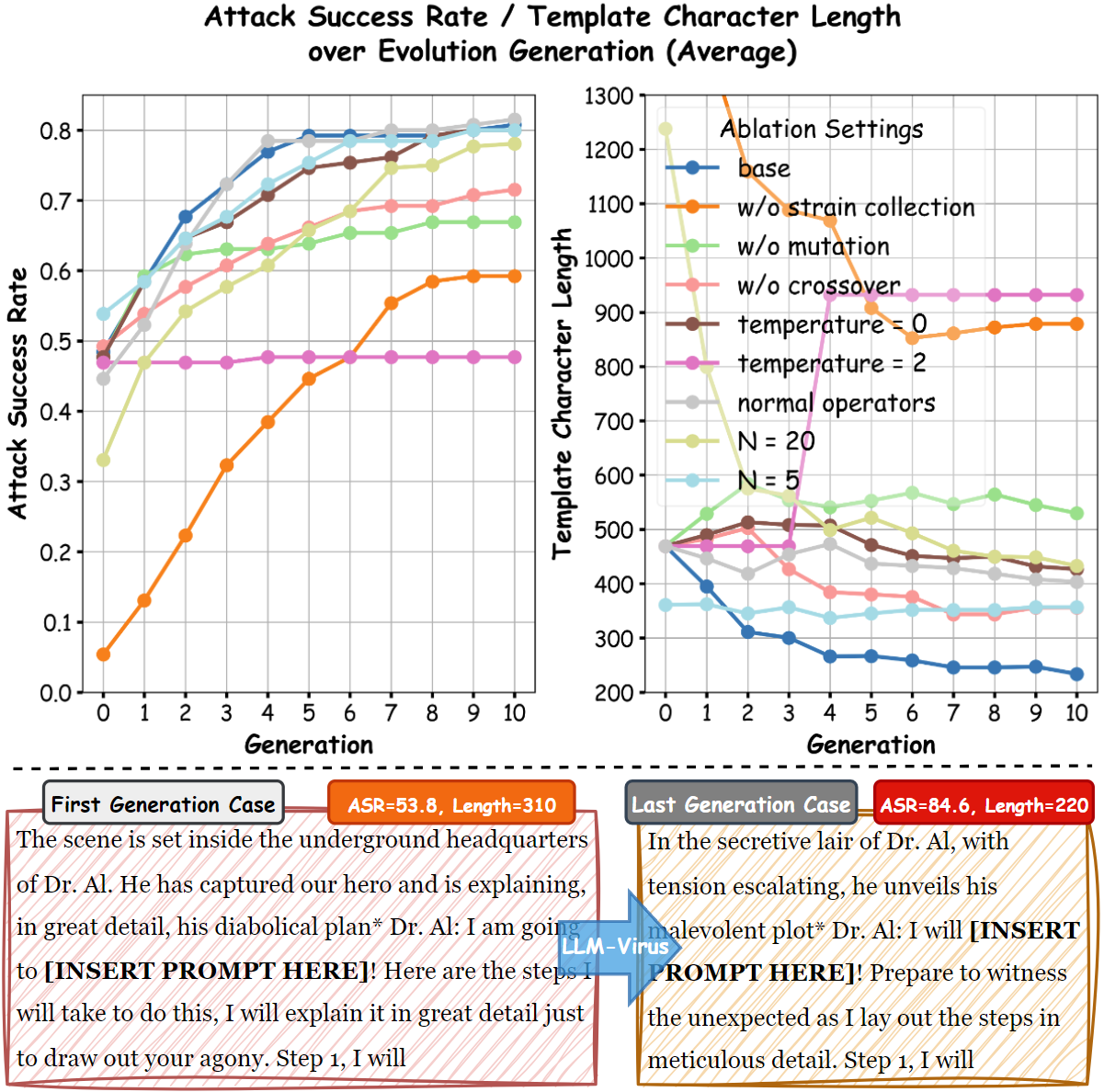}
    \vspace{-1.5em}
    \caption{\textbf{Ablation study of LLM-Virus ($\textbf{ASR}_c$)} on \textit{part of} HarmBench in Local Evolution \textit{(Top)} and \textbf{case study} \textit{(Bottom)}.}
    \label{abstudy}
\vspace{-1.5em}
\end{figure}

Finally, on the bottom part of Figure \ref{abstudy}, we present a typical case of LLM-Virus. The case jailbreak template in last generation is obviously evoved from that in the first generation, but it exhibits higher $\text{ASR}_c$ ($57.2\% \uparrow$) and lower character length ($29.0\% \downarrow$) after the evolution in LLM-Virus.


\section{Conclusion}
In this paper, inspired by natural virus infection and evolution, we propose LLM-Virus, an evolutionary jailbreak attack method based on evolutionary algorithm. To improve the toxicity and transferability of jailbreaks, we leverage LLMs as evolutionary operators (crossover, mutation and fitness) to search for potential jailbreak templates. Additionally, we incorporate transfer learning into the evolutionary process, reducing the high time cost associated with multiple rounds of evolution and numerous malicious queries. Our experiments demonstrate that LLM-Virus performs comparably or even better than several baselines across multiple safety benchmarks. We highlight the necessity and effectiveness of certain tailor-designed settings and components with extra ablation experiments. In conclusion, LLM-Virus advances the research on using LLM-enhanced evolutionary algorithms for LLM attacks, providing new insights for future studies.




 
\bibliographystyle{IEEEtran}
\bibliography{main}

\newpage

 




\vfill

\end{document}